\documentclass{aa}  

\usepackage{graphicx}
\usepackage{txfonts}
\usepackage{natbib}
\bibpunct{(}{)}{;}{a}{}{,}
\begin{document}

   \title{Star-disk interaction in classical T Tauri stars revealed using wavelet analysis}

   \author{J. L\'opez-Santiago
          \inst{1}
          \and
          I. Crespo-Chac\'on
          \inst{2}
          \and
          E. Flaccomio
          \inst{3}
          \and 
          S. Sciortino
          \inst{3}
          \and
          G. Micela
          \inst{3}
          \and 
          F. Reale
          \inst{3,4}
          }

   \institute{Dpto. de Astrof\'isica y Cencias de la Atm\'osfera, 
                  Universidad Complutense de Madrid, E-28040 Madrid, Spain\\
              \email{jalopezs@ucm.es}
         \and
                 CEDEX, E-28001 Madrid, Spain
         \and
                 INAF-Osservatorio Astronomico di Palermo, 
                 Piazza del Parlamento 1, I-90134 Palermo, Italy
         \and
                 Dipartimento di Fisica e Chimica, Universit\`a di Palermo, 
                 Piazza del Parlamento 1, I-90134 Palermo, Italy
                 }

   \date{Received ...; accepted ...}

 
  \abstract
   {The extension of the corona of classical T Tauri stars (CTTS) is under discussion. 
    The standard model of magnetic configuration of CTTS predicts that coronal magnetic 
    flux tubes connect the stellar atmosphere to the inner region of the disk. However, 
    differential rotation may disrupt these long loops. The results from Hydrodynamic 
    modeling of {X-ray flares  observed} in CTTS confirming the star-disk connection 
    hypothesis are still controversial. Some authors suggest the presence of the accretion 
    disk prevent the stellar corona to extent beyond the co-rotation radius, {while others 
    simply are not confident with the methods used to derive loop lengths.}
    }
   {We use independent procedures to determine the length of flaring loops in 
   stars of the Orion Nebula Cluster previously analyzed using Hydrodynamic models. 
   Our aim is to disentangle between the too scenarios proposed.
   }
   {We present a different approach to determine the length of flaring loops 
   based on the oscillatory nature of the loops after strong flares. We use wavelet 
   tools to reveal oscillations during several flares. The subsequent analysis of such 
   oscillations is settle on the Physics of coronal seismology. 
   }
   {{Our results likely confirm the large extension of the corona of CTTS and the hypothesis of 
    star-disk magnetic interaction in at least three CTTS of the Orion Nebula Cluster.}
   }
   {Analyzing oscillations in flaring events is a powerful tool to determine the physical
    characteristics of magnetic loops in coronae in stars other than the Sun. The results
    presented in this work confirm the star-disk magnetic connection in CTTS.}


   \maketitle
%

\section{Introduction}
\label{intro}

Classical T Tauri stars  (CTTS) and other young stellar objects are characterized by 
accreting gas from a surrounding gaseous disk. Accretion takes place 
through a \emph{funnel} that connects the star to the inner region of the disk 
\citep[e.g.][]{har98}. \citet{koe91} was the first to propose the scenario of magnetic 
driven accretion in CTTS. In such scenario, the star accretes material from the disk  
following stellar magnetic field lines connected to the disk \citep[see Figure~12 in][]{cam90}. 
A similar scenario was previously proposed for accreting neutron stars by \citet{gho78}. 
 
Some mechanisms of this star-disk interaction are still to be fully understood. Due to 
differential rotation between the star and the disk, magnetic connection may be disrupted. 
Magnetohydrodynamic (MHD) simulations predict the poloidal field connecting the disk to the star is 
wrapped up when their angular velocities differ substantially \citep{lov95}. As a consequence, magnetic loops  
inflate rapidly \citep{goo97}. This expansion of poloidal field yields a {magnetic configuration with
three components:} a closed stellar magnetosphere, a region of open field connected to the pole of the star 
and a region of open field connected to the disk. According to \citet{goo97}, reconnection of the open disk field 
with the open stellar field is produced naturally. A byproduct of the star-disk interaction is the occurrence 
of flaring events in magnetic funnels. 

\citet{orl11} performed 3D MHD modeling of {a flaring loop} 
connected to a protoplanetary disk. The authors showed that, in a time-scale of one hour after the energy is 
released close to the disk, the magnetic tube is heated ($T \sim 50$~MK) and brightens in the soft X-rays. 
The plasma inside this tube efficiently cools down in approximately one more hour. {The 
flare perturbation} propagates throughout the disk and after a few hours, {it} triggers accretion from
the opposite side of the disk onto the star following other dipolar magnetic field lines. This accretion funnel 
reaches a quasi-equilibrium state \citep[see Figure~6 in][]{orl11}. Flaring in star-disk connected magnetic 
tubes seems an efficient process to trigger accretion in T Tauri stars.

With a length $L \sim 10^{12}$~cm, the size of the hot-plasma tube created after the flare in the simulation 
by \citet{orl11} is comparable to that observed by \citet{fav05} in stars of the \emph{Chandra Orion 
Ultradeep Project} \citep[COUP;][]{get05}.
%
\citet{fav05} found that some stars of their sample exhibited long flaring loops ($> 5-10$~R$_\star$). 
Such loop lengths were compatible with the scenario of star-disk magnetic 
connection. Later, \citet{get08a} determined lower values for the loop lengths of some of those 
stars. \citet{get08a} also compared loop lengths with co-rotation radii in the COUP sample and 
obtained that X-ray coronal extent in fast-rotating, diskless stars can significantly exceed the 
Keplerian co-rotation radius, whereas X-ray loop sizes in accreting systems do not exceed it 
\citep{get08b}. They interpreted this result as an evidence that the accretion disk truncates 
the stellar magnetosphere.
However, the apparent contradiction between \citet{fav05} and \citet{get08a} in some stars arises 
from the use of different values for the slope of the decay phase in the density-temperature diagram 
of each flaring event. This characteristic of the flare is used later in both works to determine the length 
of those loops following \citet{rea97}. 

The main criticism given in the literature for using HD models {to determine loop lengths 
is that the authors usually assume} the flare occurs in a single loop instead of an arcade. 
\citet{rea07} showed that this assumption is reasonable during the first phases of the event. 
{In that work, the author obtained a diagnostic toolset to determine the loop length from 
characteristics of the rise phase of the flare. As this rise phase is typically very quick, analyzing it 
uses to be difficult due to the lack of photon statistics \citep[although see][]{lop10}.}
{As an alternative to HD modeling,} other authors apply scaling laws to determine the 
physical characteristics of stellar flares \citep[see][and references therein]{gud04}. Typically, scaling laws 
predict short loops with strong magnetic fields ($\sim 1$~KG) while single-loop HD models 
predict longer loops with weak magnetic fields ($\sim 100$~G), for the same flare. An independent 
measure of the loop lengths would help in deepening into this problem.


In recent years, oscillatory patterns have been detected in {solar flares \citep[e.g.][]{nak99,asw99}.} 
These oscillations have been interpreted as fundamental modes of MHD oscillations in coronal magnetic loops
\citep[see][and references therein]{ste12}. A variety of oscillatory phenomena have been directly observed in 
solar flaring loops at different wavelengths \citep[e.g.][]{nak01,sri08,nak09,ver09}. {Global waves traveling in 
the solar corona have been also observed \cite[see][for a review]{liu14}.} Instead, very few 
{oscillations} have been detected in flares on other stars. {Since the latter are spatially-unresolved, 
one must look for indirect signatures in {their light curve.} As a result, only oscillation modes causing changes 
in density inside the magnetic loop can be detected in stars other than the Sun.}

\citet{mit05} reported the first detection {in X-rays} of a loop oscillation in a star. The star was the young M dwarf 
AT~Mic and the oscillation was interpreted as a standing magneto-acoustic wave produced by a longitudinal, 
slow-mode wave. The physical model was based on that proposed by \citet{zai89}: {the flaring loop oscillates 
due to a centrifugal force caused by the filling in of the loop by chromospheric material. 
Previously to \citet{mit05},} \citet{mat03} assumed the same model to study the oscillation observed 
during the decay phase of a {white-light flare observed in the U-band} in the chromospherically active binary II~Peg. 
\citet{mat06} analyzed an optical 
oscillation observed in EQ~Peg~B, another M dwarf star. In that case, the short oscillation observed in the U-band 
was interpreted as a fast-MHD wave. In particular, the authors proposed the sausage mode was responsible for 
the oscillation in the light curve. More recently, \citet{pan09} found evidences of a fast-kink mode in the X-ray 
oscillations detected in $\tau$~Boo, a G/K binary system. {However, kink modes do not cause 
changes in density and they should not be detected in the light curve of flaring stars \citep[see discussion in][]{ste12}.}

In a study of UV light curves of M dwarfs, \citet{wel06} demonstrated that oscillations are frequent during stellar flares. 
However, these oscillations are effectively damped by thermal conduction {\citep{ofm02,sel05,jes12}.} As a result, only a few 
periods (typically 3-4) are observed. Numerical simulations of magneto-acoustic waves in solar coronal loops
show that the initial impulse mainly triggers the fundamental mode and its first harmonic \citep{sel05}. Their
study provides the longitudinal density stratification in the loop \citep[][and references therein]{sri13}. 

We aim at analyzing oscillatory patterns in the young stars of the {Orion Nebula Cluster observed by the 
COUP}, in order to determine the properties of their flaring loops, including loop length. We will compare 
our results with those obtained by \citet{fav05} and \citet{get08a} using HD models. We notice that
this method to determine loop lengths is very robust since it is based only on basic Physics laws. 
{We do not make \emph{a priori} assumptions on the aspect of the loop or any other 
characteristic. If an oscillation is revealed in the light curve of a star, it must come from a single loop
as the ignition of different loops in an arcade is produced at distinct times. In other words, the oscillation 
of different loops in an arcade is not coherent and if more than one loop oscillates even at the same frequency 
and intensity, the signal of the distinct oscillations would cancel each other and it would 
not be detected in the light curve of the star.} 
The light curves of the stars are analyzed using wavelet methods. 
   

\section{Wavelet analysis of the COUP sample}
\label{method}

For our work, we used wavelet analysis tools to detect any oscillation pattern.
The wavelet analysis is a modified version of the Fourier analysis adapted
for detecting quasi-periodic signals. The light curve is transformed {from the time domain} 
into the frequency domain by convolving the original signal with a \emph{mother} function (a \emph{wavelet}). 
In a Fourier transform, the mother function is a combination of sines and cosines that are infinite in the time 
domain. As a result, {by using a Fourier transform, one cannot know the exact moment the oscillation 
was triggered or whether it remained for a long lapse of time or it was rapidly damped.} Instead, using 
particular mother functions, {this kind of information} can be {preserved}. Following \citet{tor98}, we used a 
Morlet function as the wavelet, which has been previously used with success to reveal oscillations during 
{solar flares \citep[e.g.][]{ofm00,ban00,dem02,ver04,sri08}} and stellar flares \citep[e.g.][]{mit05,wel06,mat06,gom13,pil14}: 

\begin{equation}
\phi(\eta) = \pi^{-\frac{1}{4}} e^{i\omega_0\eta} e^{-\frac{\eta^2}{2}}
\end{equation}

where $\eta$ is a non-dimensional time parameter and the non-dimensional frequency ($\omega_0$)
is set equal to 6 to approximately satisfy the admissibility condition, {i.e. the Morlet function with 
$\omega_0 = 6$ integrates close to zero or, in other words, it permits a reconstruction of the original signal.}

The square of the wavelet transform of the signal is the (wavelet) power spectrum and, for a time-dependent 
mother function such as the Morlet one, it is two-dimensional (one dimension for frequency and one dimension for time). 
The peaks showing a high confidence level in this power spectrum are likely related to oscillation patterns or any \emph{disturbance} 
in the time series (e.g. a flaring event). In particular, oscillation patterns maintained in time are detected as 
an extended peak in the power spectrum. To investigate the properties of these patterns, the light curve can be 
reconstructed by filtering the power spectrum at those periods with high significance \citep{tor98}. 

For the case of a Morlet function, confidence levels can be derived bearing in mind that the power spectrum 
is $\chi^2$ distributed \citep{ge07}. In particular, for a given background spectrum $P_k$, the power spectrum of the
wavelet transform ($W_n$) satisfies the condition: 

\begin{equation}
\frac{\vert W_n \vert^2}{\sigma^2}  \longrightarrow \frac{1}{2} P_k \chi^2_2
\label{eqn2}
\end{equation}

where $\sigma^2$ is the variance of the signal in the time domain and $\chi^2_2$ is the two dimensional $\chi^2$ distribution. 
Here, the $\longrightarrow$ symbol means ``is distributed as''. $\vert W_n \vert^2$ is the power spectrum of the 
wavelet transform of the signal. We notice that, for general white noise background, $P_k = 1$.

{To determine confidence levels for the features detected in the wavelet power spectra, 
we assumed the background spectrum is red noise instead of white noise, which is typically 
used in the literature. We note that the global shape of the flare must be subtracted to the the observed 
light curve first to determine robust confidence levels. We performed this subtraction 
by smoothing the flare light curves by a moving average and polynomial fitting.}

{Finally, the original time series can be reconstructed later by filtering the power spectrum 
at periods with high significance using either deconvolution or the inverse filter. 
Nevertheless, in our work we follow \citep{tor98} and reconstruct the original light curve
as the sum of the real part of the wavelet over all the scales.}

\section{A simple model for the triggering of oscillations during a flare}
\label{model}

Let we consider the oscillation is triggered by an increment in the magnetic pressure that produces a
change in the volume of the magnetic tube. This scenario gives rise to sausage modes in the perturbed 
plasma. By the conservation of linear momentum, the force over a unit of volume in the tube is

\begin{equation}
\vec{F} = \vec{J} \times \vec{B} - \vec{\nabla} p 
\label{eqmagforce}
\end{equation}

where $\vec{J}$ is the current density and $p$ is the plasma pressure. From the Amp\`ere's law  
applied to a plasma

\begin{equation}
\vec{J} \times \vec{B} = -\vec{\nabla}\left(\frac{B^2}{2\mu_0}\right) + (\vec{B} \cdot \vec{\nabla}) \vec{B}/\mu_0
\label{eqampere}
\end{equation}

Combining Eqs.~\ref{eqmagforce} and \ref{eqampere}, 

\begin{equation}
\vec{F} = -\vec{\nabla}p - \vec{\nabla}\left(\frac{B^2}{2\mu_0}\right) + \frac{1}{\mu_0} (\vec{B} \cdot \vec{\nabla}) \vec{B}
\label{eqampforce}
\end{equation}

The second term in Equation~\ref{eqampforce} describes the magnetic energy density and acts as the magnetic 
pressure

\begin{equation}
p_m = \frac{B^2}{2\mu_0}
\label{eqmagpress}
\end{equation}

The third term in Equation~\ref{eqampforce} is the magnetic stress. If we consider an unidirectional magnetic field, 
$\vec{\nabla} \cdot \vec{B} = 0$. Hence, $(\vec{B} \cdot \vec{\nabla}) \vec{B} = 0$ and, assuming static equilibrium, 
$p = p_m$. 

Let $\sigma = B_\mathrm{max}/B_\mathrm{min}$ be the magnetic mirror ratio of the loop, with B$_\mathrm{max}$
the magnetic field strength at the loop base and B$_\mathrm{min}$ the magnetic field strength at the top of the 
loop (loop apex). Assuming B$_\mathrm{max}$ constant, $\Delta \sigma / \sigma = 
\Delta B_\mathrm{min}/B_\mathrm{min}$. Therefore, for static equilibrium,

\begin{equation}
\frac{\Delta B_\mathrm{min}}{B_\mathrm{min}} = \Delta p \frac{\mu_0}{B_\mathrm{min}^2} \approx \frac{4 \pi n_e k_B T}{B_\mathrm{min}^2} = \beta/2
\label{eqdeltab}
\end{equation}

Here, $n_e$ and T are the plasma density and temperature, respectively and $\beta$ is the ratio of the plasma pressure 
($p \approx n_e k_B T$) to the magnetic pressure ($p_m = B_\mathrm{min}^2/2\mu_0$) at the loop apex. 

By the conservation of the magnetic flux, the change in the volume of the loop is inversely proportional to the change 
in the magnetic field strength

\begin{equation}
\frac{\Delta V}{V} = \frac{\Delta n_e}{n_e} = -\frac{\Delta B}{B}
\label{eqdeltav}
\end{equation}

If the plasma is considered optically thin, the bremsstrahlung flux may be expressed as 

\begin{equation}
I \propto \frac{n_e^2}{\sqrt T} V
\label{eqintensity}
\end{equation}

and, combining Eqs.~\ref{eqdeltab} to \ref{eqintensity}

\begin{equation}
\frac{\Delta I}{I} \approx \frac{4 \pi n_e k_B T}{B_\mathrm{min}^2}
\label{eqdeltai}
\end{equation}

As a result, for sausage-type oscillations, the magnetic field strength at the top of the loop can be inferred from 
the amplitude of the oscillation, if the characteristics of the plasma are previously determined from observations. 

{Alternative scenarios for the production of quasi-periodic pulsations in flares (slow magneto-acoustic modes, 
fast modes, Kelvin--Helmholtz instabilities, etc.) have been proposed in the literature \citep[see][for a review]{nak09}. 
Many of the discussed processes have negligible effects on 
the soft X-ray light curve. For example, \citet{ofm06} suggest that super-Alfv\'enic beams in the vicinity of the 
reconnection region lead to the excitation of the oscillations through the coupling of tearing mode and 
Kelvin--Helmholtz instabilities. However, they also note that the modulation is not observed in thermal 
emission. In addition, fast magneto-acoustic modes are highly dispersive and have typical periods 
quite shorter than those observed in our sample.}

\begin{figure*}[!t]
   \centering
   \includegraphics[width=\columnwidth]{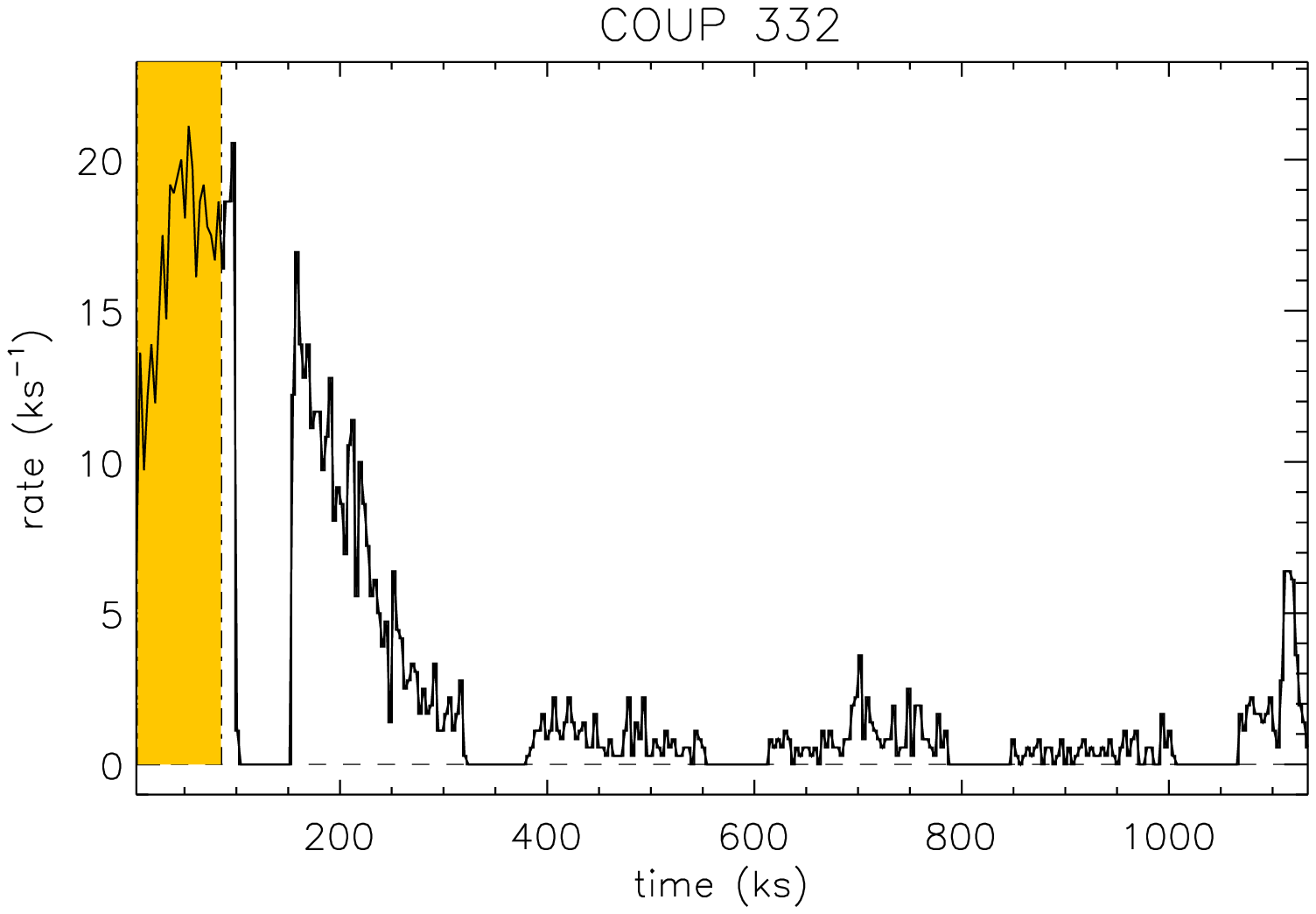}
   \includegraphics[width=\columnwidth]{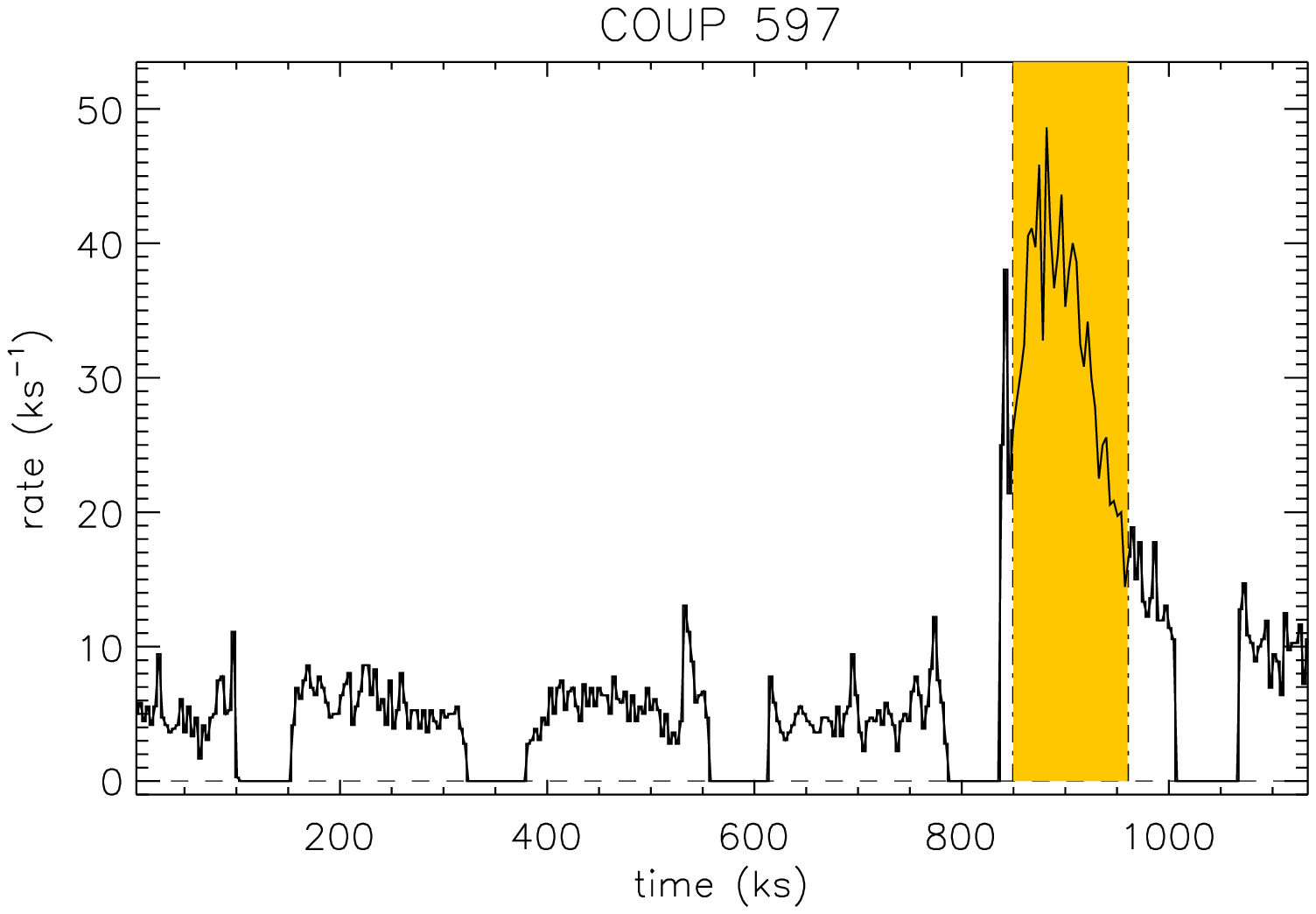} \\
   \vspace{0.5cm}
   \includegraphics[width=\columnwidth]{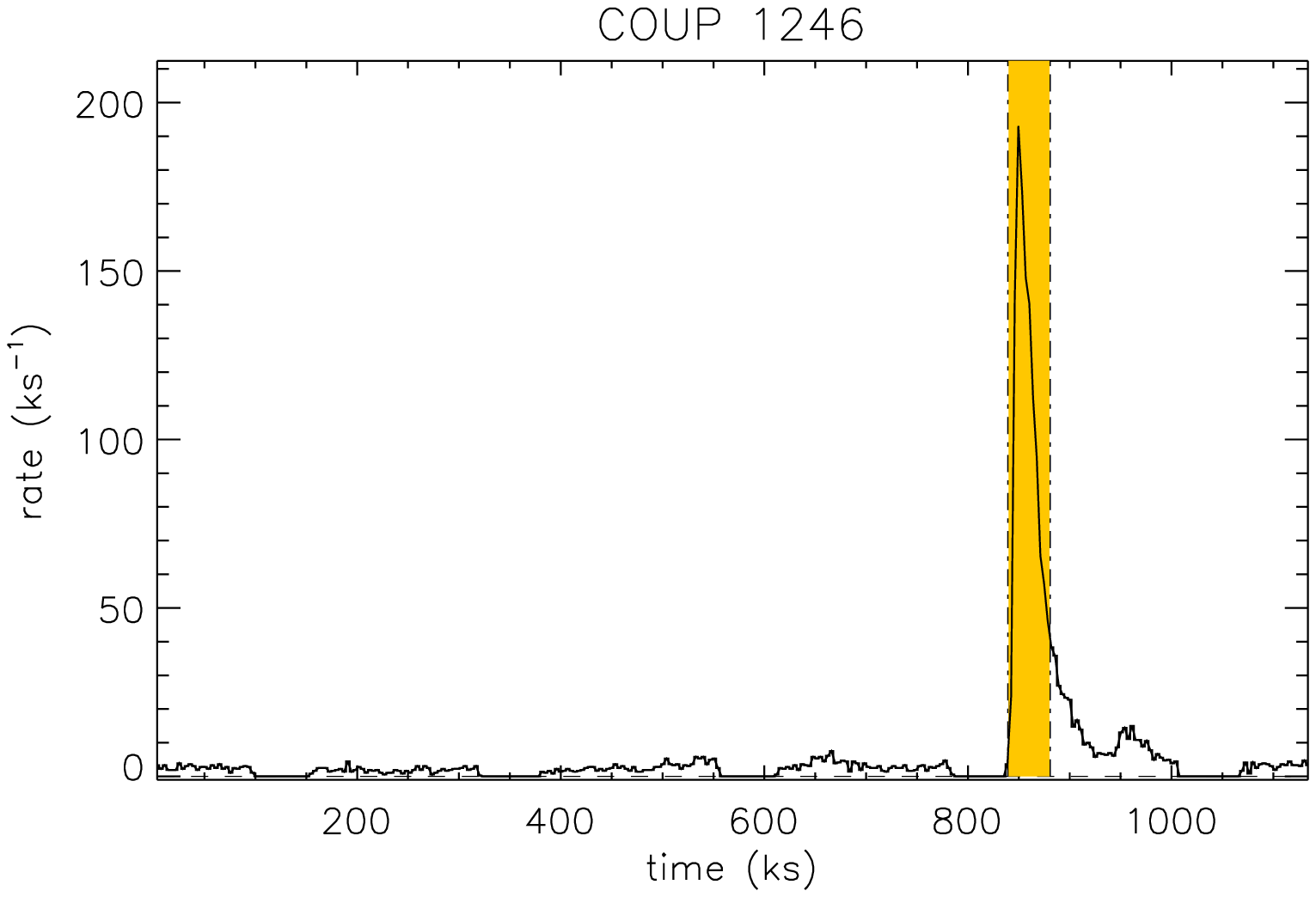}
   \includegraphics[width=\columnwidth]{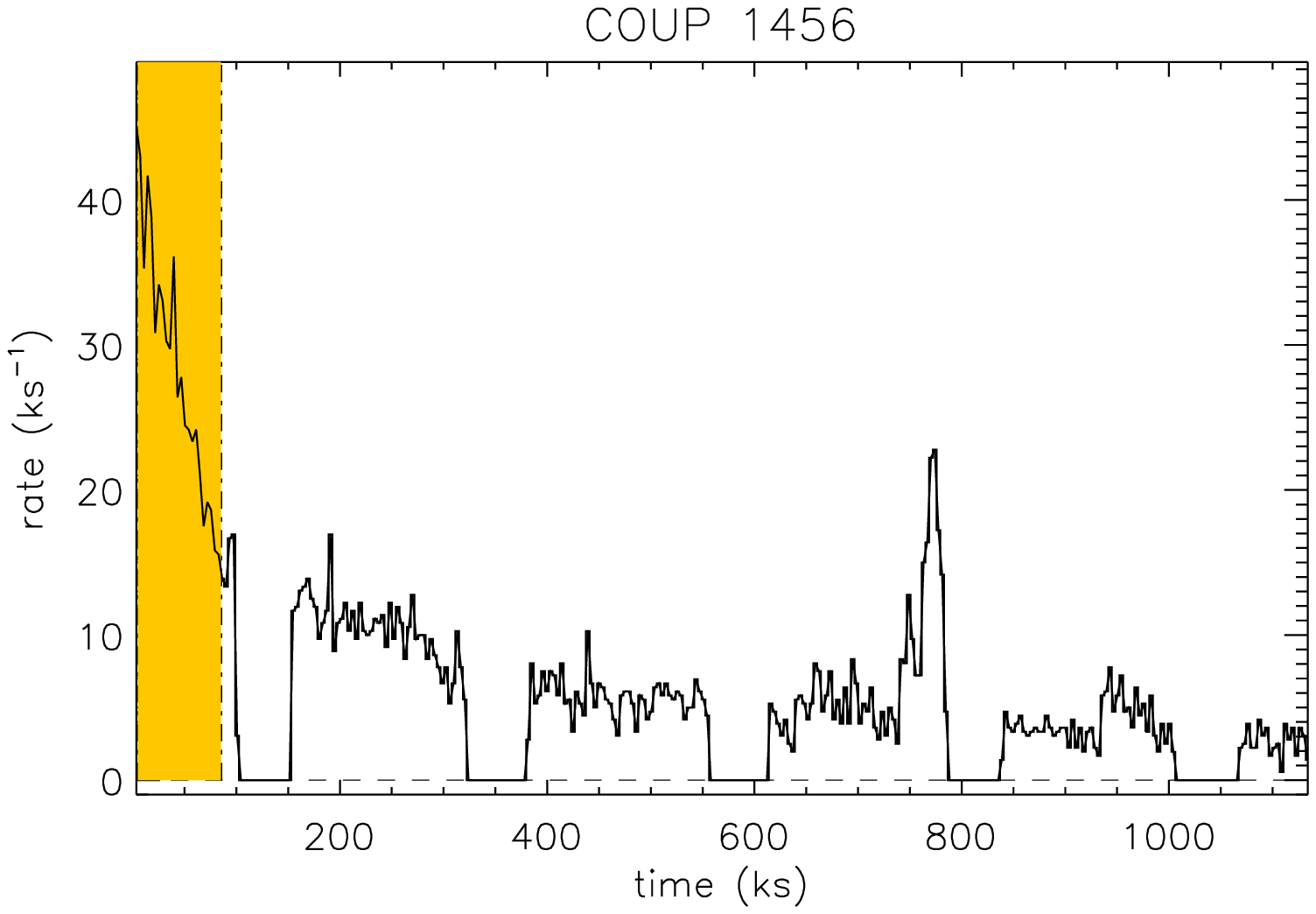} \\
   \vspace{0.5cm}
   \includegraphics[width=\columnwidth]{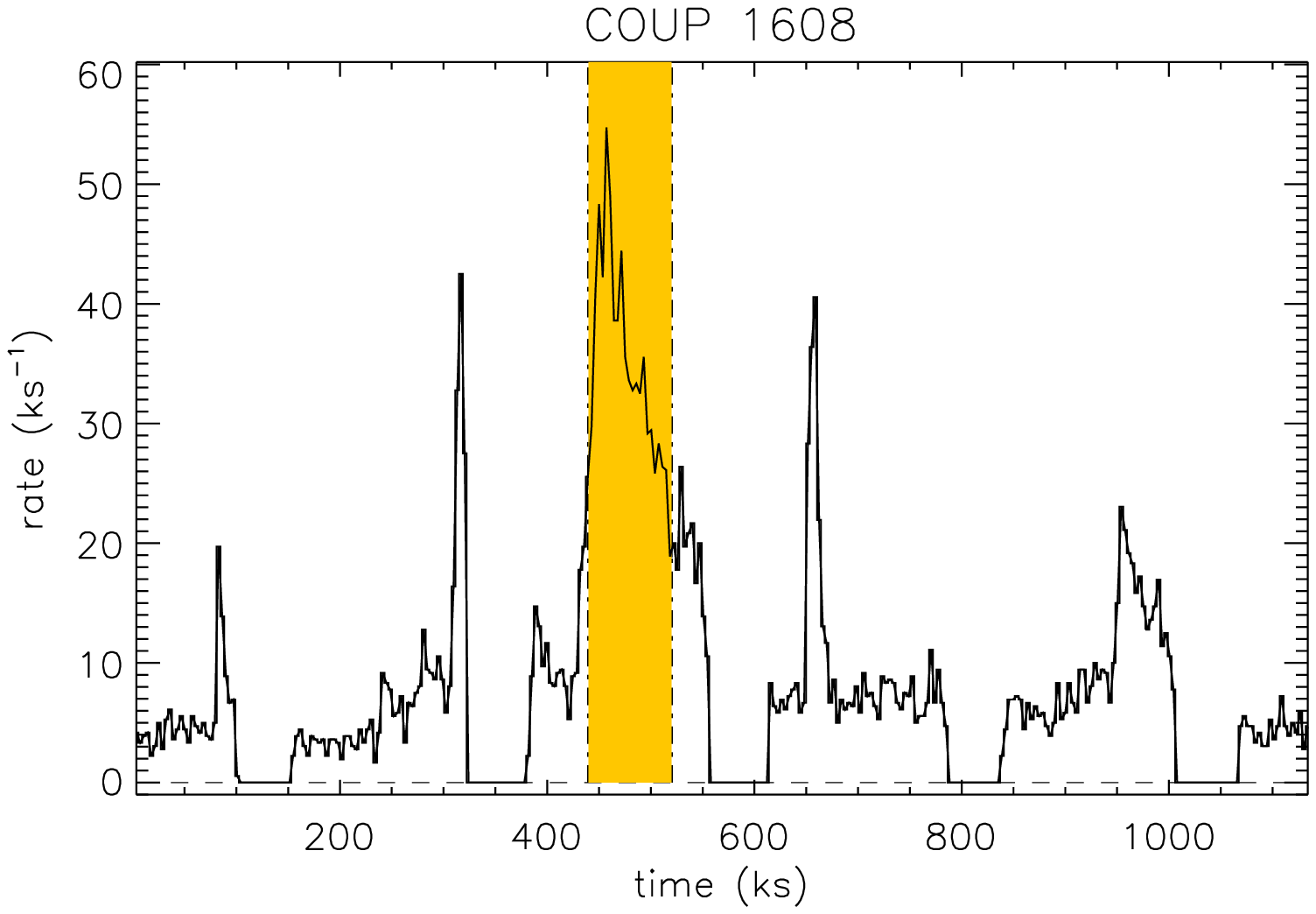} 
   \caption{Observed light curves of the studied COUP sources. 
                 Time binning is 1~h. The shadowed area marks the time 
                 interval studied in this work. 
   \label{curves}}
\end{figure*}

\section{Discussion of individual flares}
\label{results}

From the sample of COUP sources showing large flares, we selected those sources studied by \citet{fav05} and 
\citet{get08b} with the aim of comparing results. Each source was studied independently because the
behavior of the flare is different in each star and, thus, the source and/or characteristics of the oscillations
can differ. We focussed only on the light curves with significant oscillatory patterns. {Figure~\ref{curves} 
shows the light curve observed for each star during the COUP observations. Time binning is 1~h.}

In the following, we present results for individual sources showing different behaviors. {A summary of the 
results compared with those of \citet{fav05} and \citet{get08a} is shown in Table~\ref{table}.} 
{For completeness, we compare the results obtained assuming sausage-type oscillations 
(MHD scenario) with those of acoustic waves (HD scenario). Details on this comparison are given 
for each star.}

\subsection{COUP 332}
\label{sec:cou332}

The source COUP 332 (2MASS J05350934-0521415) underwent a long-duration flare ($> 300$~ks) during the  
observations. \citet{fav05} derived a peak temperature $T_\mathrm{peak} = 113$~MK and a very low electronic 
density $n_e = 0.05 \times 10^{10}$~cm$^{-3}$. Applying \citet{rea97}, the authors determined a loop length 
$L \sim 7 \times 10^{12}$~cm. This value is considerably higher than the one derived by \citet{get08a}, who 
obtained $L \sim 0.15 - 1.7 \times 10^{12}$~cm. 
{\citet{aar10} suggest a star-disk connection scenario in COUP~332, from the dust destruction radius 
estimated by them (see Table~\ref{table}) and the loop length derived by \citet{fav05}.}

\begin{figure}[!t]
   \centering
   \includegraphics[width=\columnwidth]{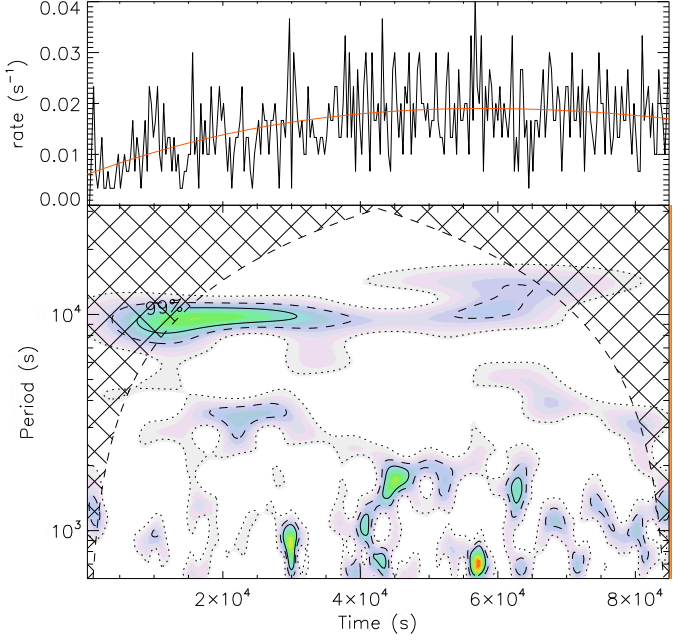} 
   \caption{Wavelet power spectrum of a flare of source COUP~332 in the 
                 bottom panel. Contours are for confidence levels 67\% (dotted line), 95\% 
                 (dashed line) and 99\% (continuous line). {The hatched area is the cone of 
                 influence (COI), the region of the wavelet power spectrum in which edge effects 
                 become important \citep{tor98}.} Top panel shows the observed
                 light curve. Overplotted as a continuous line is the flare's global shape. 
                 Time binning is 300~s. 
   \label{coup332}}
\end{figure}

The COUP observations are divided into periods of approximately 100 to 160~ks interrupted 
by 50~ks gaps. Together with the long duration of the flare of COUP~332, this fact causes a 
gap during the decay phase. We analyzed the first part of the flare, consisting on the rise phase 
and the first 50~ks of the decay, when the oscillation should be triggered. Figure~\ref{coup332}
shows the result of the wavelet analysis. The top panel of the figure is the observed light curve, 
with a moving average overplotted as a continuous line. {The bottom panel is the wavelet 
power spectrum after subtraction of the moving average to the observed light curve. 
The hatched area in the bottom panel is the cone of influence (COI), the region of the wavelet 
power spectrum in which edge effects become important \citep{tor98}. For our work, we analyze 
only features outside the COI.}

A clear feature is detected during the rise phase ($P \sim 9$~ks) that evolves during more 
than 20~ks. The sound speed inside the tube, for an adiabatic plasma with $\gamma = 5/3$,
is  $c_s = 1.16 \times 10^4 \sqrt T = 1.2 \times 10^8$~cm\,s$^{-1}$. If we assume the 
oscillation with period $P = 9 \pm 1$~ks was caused by a wave moving along the 
tube at the velocity of the sound and this is the fundamental mode, the loop length 
would be $L = 1.1 \pm 0.1 \times 10^{12}$~cm. This value is in the range given by 
\citet{get08a} and lower than that given by \citet{fav05} who likely overestimated the 
loop length. {We notice that \citet{fav05} did not determine uncertainties in the loop 
lengths and any comparison with our results should be made with caution.}

\begin{figure}[!t]
   \centering
   \includegraphics[width=\columnwidth]{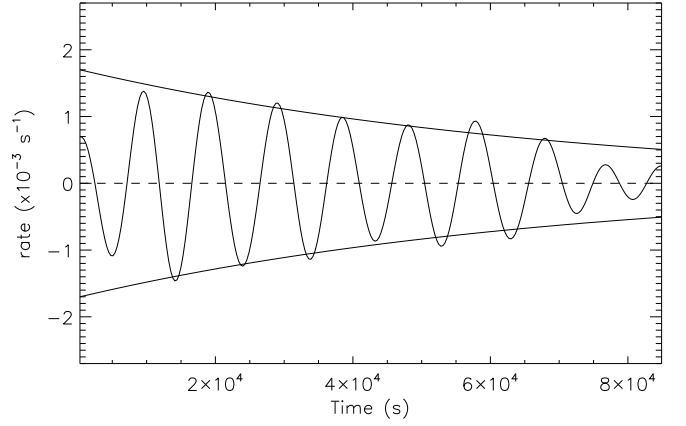}
   \caption{Reconstructed light curve of COUP~332 for period $9 \pm 1$~ks. 
                 The continuous lines are an exponential decay with e-folding time of $70$~ks.  
                 Time binning is 300~s.  
   \label{coup332rec}}
\end{figure}

From the light curve reconstruction at the period of the oscillation (see Fig.~\ref{coup332rec}), 
the intensity of the perturbation is $\Delta I / I = 0.17$. From Eq.~\ref{eqdeltai}, the magnetic field 
strength at the loop apex is $B_\mathrm{min} = 24 \pm 3$~G. This value is slightly 
higher than that given by pressure equilibrium \citep[$B = 12$~G;][]{fav05}. With 
an Alv\'en velocity $c_a \approx 2.18 \times 10^{11} B / \sqrt n_e = 2.4 \times 10^8$~cm\,s$^{-1}$,
the tube velocity for our plasma is $c_t = 1.1 \times 10^8$~cm\,s$^{-1}$. For an 
oscillation period $P = 9 \pm 1$~ks, the loop length becomes $L = 0.98 \pm 0.11 \times 10^{12}$~cm,
close to the length determined by assuming a wave moving at the velocity of the sound in the 
plasma. 

{Figure~\ref{coup332rec} also shows that the oscillation is damped. The quality factor of the 
oscillation is $Q \approx 20$ and the e-folding time is $t_d =  60 \pm 5$~ks. Comparing results with 
\citet{ofm02}, the damping of the oscillation does not seem to follow the scaling laws derived for 
thermal conduction dissipation in the Sun. The ratio $t_d / P = 6.7 \pm 1.9$ is within the range 
obtained from simulations by \citet{sel05}, where the authors considered only energy leakage into the 
photosphere as the wave damping mechanism.}

\subsection{COUP 597 (V2252 Ori)}
\label{sec:coup597}

\begin{figure}[!t]
   \centering
   \includegraphics[width=\columnwidth]{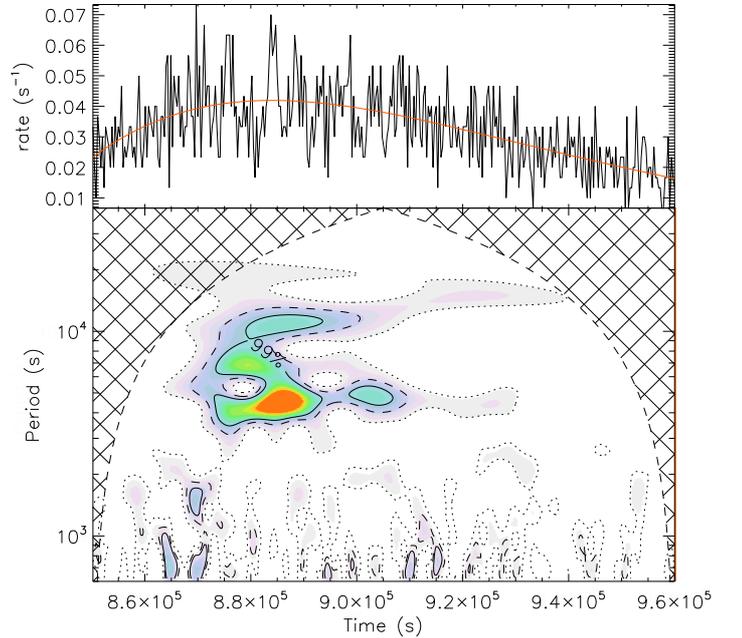} 
   \caption{Wavelet power spectrum of a flare of source COUP~597 (V2252~Ori) in the 
                 bottom panel. Top panel shows the observed light curve in the time interval of the 
                 flare. Lines are as in Fig.~\ref{coup332}. Time binning is 300~s. 
   \label{coup597}}
\end{figure}

\begin{figure}[!t]
   \centering
   \includegraphics[width=\columnwidth]{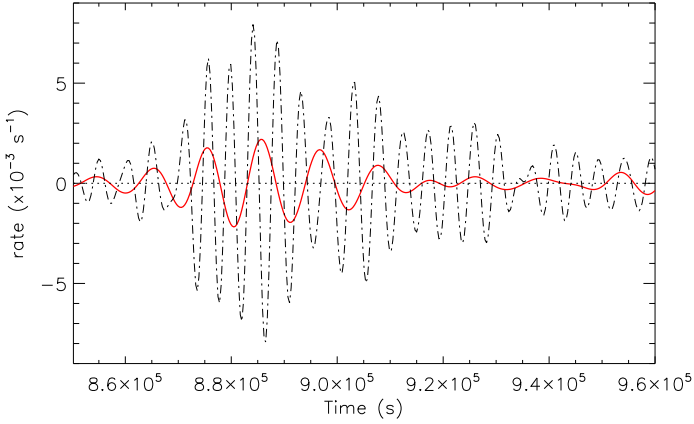} 
      \caption{Light curve reconstruction of COUP~597 (V2252~Ori) at periods 
                   $P = 4.5$~ks (continuous line) and $P = 10$~ks (dotted-dashed line). 
                   Time binning of 300~s. 
   \label{coup597rec}}
\end{figure}

Figure~\ref{coup597} shows the wavelet power spectrum of the flare detected during the COUP in V2252~Ori. 
The power spectrum of the subtracted light curve shows a significant feature extended between 4 and 12~ks. Its shape
resembles somewhat that of Prox~Cen during the flare underwent in 2009 \citep{sri13}. In that case, the authors identified 
two peaks in the power spectrum of the light curve and considered the longer period as the fundamental mode and the 
shorter period as the first harmonic. This scenario may be correct also for the flare observed in V2252~Ori. 
Figure~\ref{coup597rec} shows the reconstructed curve at periods $P \sim 4.5$~ks (dotted-dashed line) and 
$P \sim 10$~ks (continuos line). The amplitude of the $4.5$~ks feature is higher. According to \citet{sel05}, 
if we assume this feature is the first harmonic of the oscillation, the pulse of energy release should have occurred 
closer to the loop foot than to its apex.  

With the plasma parameters determined by \citet[][$T_{peak} = 87$~MK and $n_e = 5.2 \times 10^{10}$~cm$^{-3}$]{fav05},
the sound velocity inside the flaring loop is $c_s = 1.16 \times 10^4 \sqrt T = 1.1 \times 10^8$~cm\,s$^{-1}$ (assuming 
the plasma is adiabatic,  with $\gamma = 5/3$). If the oscillation was caused by an
acoustic mode and we assume the 10~ks feature is the fundamental mode of the oscillation, $L = 1.1 \pm 0.2 \times 10^{12}$~cm. 
If we consider the 4.5~ks is the fundamental mode, $L = 4.9 \pm 0.5 \times 10^{11}$~cm. Both values of the loop length 
are close, but higher than that obtained by \citet{fav05}. According to \citet{aar10}, this flaring loop is possibly anchored 
to the inner protoplanetary disk (see Table~\ref{table}). 

As discussed in Section~\ref{model}, for an optically thin plasma in static equilibrium, the magnetic field at the top of the loop 
can be determined if the amplitude of the oscillation is known (see Eq.~\ref{eqdeltai}). For our case, $\Delta I / I = 0.3 \pm 0.1$
and $B = 162 \pm 27$~G. This value is {of the order of} that determined from pressure equilibrium \citep{fav05}. 
The Alfv\'en velocity inside the loop is $c_a = \approx 2.18 \times 10^{11} B / \sqrt n_e = 1.55 \times 10^8$~cm\,s$^{-1}$ 
and the tube velocity {(the minimum velocity of the slow mode)}, 
$c_t = 8.97 \times 10^7$~cm\,s$^{-1}$. For this velocity, $L = 4.0 \pm 0.5  \times 10^{11}$~cm assuming the 4.5~ks 
feature is the actual fundamental mode, $L = 9.0 \pm 0.9  \times 10^{11}$~cm if we use the 10~ks {(similar to the value 
derived assuming acoustic waves).}

{The two oscillations detected are not very efficiently damped ($Q \approx 25$; see Figure~\ref{coup597rec}).
Contrarily to COUP~332, the damping time of these oscillations $\tau_d \sim 45~ks$ do follow the scaling law of \citet{ofm02}
for low temperature plasma ($T \sim 6$~MK). Those laws were obtained for wave dissipation by thermal conduction.} 





\subsection{COUP 1246 (V1530 Ori)}
\label{sec:coup1246}

V1530~Ori was included in \emph{Category 3} (``the inner edge of the dust disk is clearly beyond reach of the
magnetic flaring loop'') by \citet{aar10}. Consequently, the flaring loop cannot be anchored to the accretion disk. 
\citet{fav05} determined a loop length $L = 4.0 \times 10^{11}$~cm ($L/R_\star = 3.4$) and a magnetic field strength 
$B = 240$~G. The value for the loop length determined by \citet{get08a} is slightly smaller ($L \sim 2.4 \times 10^{11}$~cm)
but still the loop is long ($L/R_\star \sim 2$). 

\begin{figure}[!t]
   \centering
   \includegraphics[width=\columnwidth]{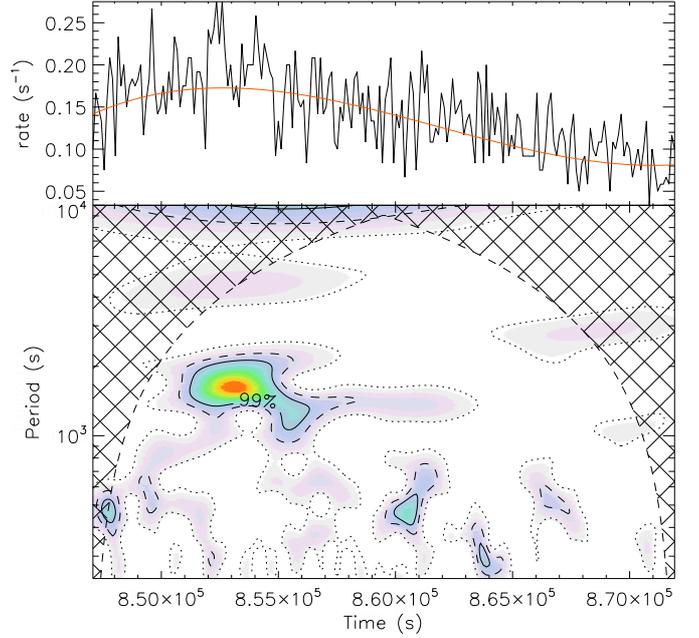}
   \caption{Wavelet power spectrum of the source COUP~1246 (V1530~Ori). Time binning 
                 is 120~s. Panels and lines are as in Figure~\ref{coup597}.
   \label{coup1246}}
\end{figure}

The wavelet power spectrum of the flare observed in V1530~Ori (Figure~\ref{coup1246}) 
shows a feature {outside the cone of influence.} Thus, we do not analyze it here. The feature at 
$P \approx 1.8$~ks is detected during the rise phase, close to the flare maximum. From the 
peak temperature $T = 270$~MK determined by \citet{fav05}, the sound speed of the plasma inside 
the magnetic tube is $c_s =  1.9 \times 10^8$~cm\,s$^{-1}$. An acoustic wave would travel $3.4 \times 10^{11}$~cm 
in 1.8~ks. This value is close to the loop length determined by \citet{fav05}. 

\begin{figure}[!t]
   \centering
   \includegraphics[width=\columnwidth]{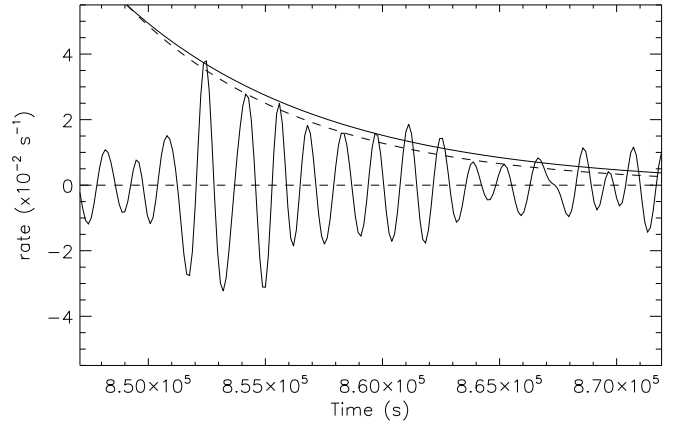}
   \caption{Reconstructed light curve of COUP~1246 (V1530~Ori) for periods $1.0-2.0$~ks. 
                 Time binning is 120~s. Continuous line is a exponential decay with e-folding time 
                 of 8.5~ks (dashed line is for e-folding time of 7.5~ks).  
   \label{coup1246rec}}
\end{figure}

The reconstructed light curve for the period range $1.0~\mathrm{ks} < P < 2.0$~ks is shown in 
Figure~\ref{coup1246rec}. The e-folding time of the oscillation is $8.5 \pm 1.0$~ks. 
This value does not follow the scaling laws for damping in solar flares treated in 
\citet{ofm02} and \citet{mar05}. In those works, damping was driven by thermal conduction processes. 
The long decay time obtained in this oscillation suggests the damping in this flare was not caused 
by thermal conduction {but by leakage into the photosphere \citep[see][]{sel05}.}

The intensity of the oscillation is $\Delta I / I = 0.2 \pm 0.04$. From Eq.~\ref{eqdeltai}, the magnetic field 
strength at the loop apex is $B_\mathrm{min} = 362 \pm 45$~G, considerably higher than the 
value derived from pressure equilibrium \citep[$B = 240$~G;][]{fav05}. With this magnetic field, the 
Alfv\'en velocity inside the tube is $c_A =  3.3 \times 10^8$~cm\,s$^{-1}$ 
\citep[assuming $n_e = 5.6 \times 10^{10}$~cm$^{-3}$;][]{fav05}, for a tube velocity 
$c_t =  1.7 \times 10^8$~cm\,s$^{-1}$. Using a period $P = 1.8 \pm 0.2$~ks for the fundamental mode
of oscillations, we derive a length loop $L = 3.4 \pm 0.4 \times 10^{11}$~cm, as for the case of an acoustic 
wave. 

\subsection{COUP 1456}
\label{sec:coup1456}

\begin{figure}[!t]
   \centering
   \includegraphics[width=\columnwidth]{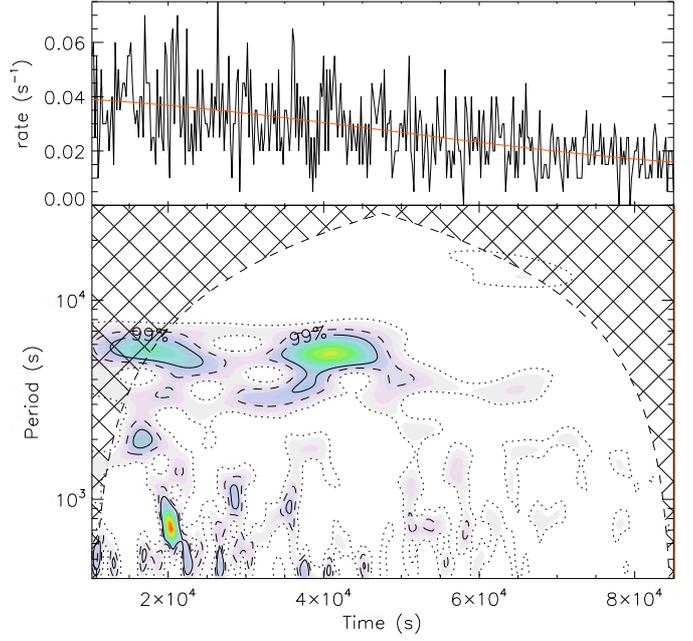}
   \caption{Light curve and wavelet power spectrum of the source COUP~1456. 
                 Time binning is 200~s. Lines are as in Figure~\ref{coup332}.
   \label{coup1456}}
\end{figure}

During the COUP observations, this source showed a continuous decrease in its count rate during more than 
500~ks, likely produced by the decay of a large flaring event. This flare was not studied by \citet{fav05} or 
\citet{get08a}, most likely because the flare maximum took place before the observation started. 
In the wavelet power spectrum of the first 90~ks of the observation (Figure~\ref{coup1456}) the source shows 
a double peak feature at $\sim 6$~ks. Before it, a peak at $\sim 3.0$~ks is detected. The peak at the highest 
frequency may be related to the first harmonic of the oscillation and the feature with the longest period 
to the fundamental mode. The detection of the first harmonic before the detection of the fundamental mode 
can be related to the location of the initial pulse (energy input) in the loop \citep{sel05}. The first harmonic 
is excited when the initial pulse is launched close to the top of the loop. This harmonic is more sensitive to 
the inhomogeneities of the medium and, hence, it is more efficiently damped. On contrast, the fundamental 
wave tends to occupy the entire loop over its length and it is damped less efficiently. 

We may assume the temperature at the beginning of the decay phase is $T = 20-40$~MK 
taking the results of \citet{get08a} for other sources into account. For a 20~MK plasma, the sound velocity 
in the medium is $c_s = 7.4 \times 10^7$~cm\,s$^{-1}$. For the fundamental wave ($P = 6 \pm 0.5$~ks) 
moving at the sound velocity, $L = 4.4 \pm 0.4 \times 10^{11}$~cm, similar to other loop lengths determined 
in previous sections for other flaring loops. 

\subsection{COUP 1608 (OW Ori)}
\label{sec:coup1608}

\citet{get08b} concluded that OW~Ori is the only star in the COUP sample where the flare took place in a
loop connecting the star with the disk. Actually, the authors assumed that the disk of the stars in their 
sample were all truncated at the stellar co-rotation radius. \citet{aar10} showed that this assumption is 
wrong in some cases. Nevertheless, OW~Ori remains the most promising case to show a flare in 
a star-disk connected magnetic tube because the length derived in the literature by distinct authors 
is always above the co-rotation radius (see Table~\ref{table}). \citet{get08a} determined a loop length of 
$\sim 8 \times 10^{11}$~cm, similar to the value determined by \citet{fav05}. This length corresponds to 
6.7~R$_\star$, which is longer than the co-rotation radius. This result strongly suggests the energy release 
occurred in a magnetic tube connecting the star with the accretion disk. 

\begin{figure}[!t]
   \centering
   \includegraphics[width=\columnwidth]{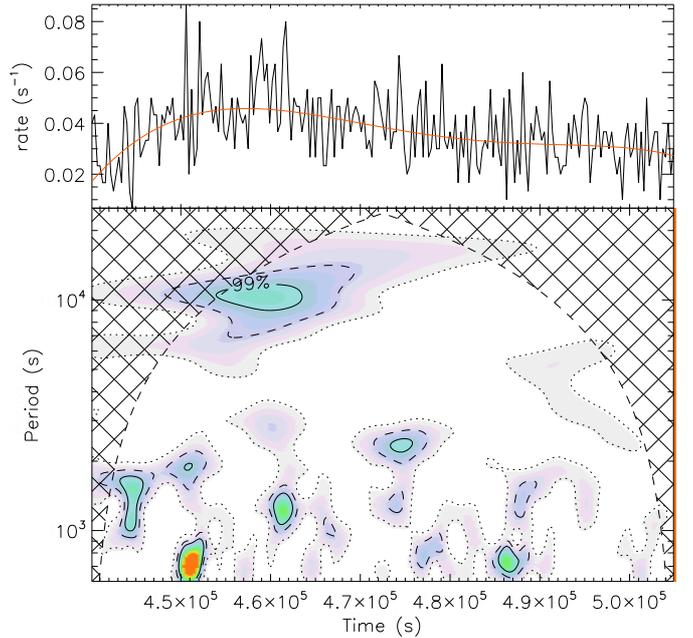}
   \caption{Wavelet power spectrum of the source 
                 COUP~1608 (OW Ori). Lines and panels are as in Figure~\ref{coup332}.
                 Time binning is 300~s. 
   \label{coup1608}}
\end{figure}

{Figure~\ref{coup1608} shows the wavelet power spectrum} of the flare light curve of OW~Ori in the COUP. 
The global shape of the flare was subtracted from the light curve by smoothing the observed curve with 
a moving average. The feature at $P \approx 10$~ks is clearly identified. It starts just before  
the flare maximum. According to the parameters determined in \citet{fav05} for this flare, 
$n_e = 10^{10}$~cm$^{-3}$ and $T_{peak} = 258$~MK. The sound velocity inside the tube is, 
then, $c_s = 1.8 \times 10^8$~cm\,s$^{-1}$. Assuming the feature at $P = 10$~ks is the fundamental 
mode of an oscillation propagating along the magnetic tube, $L = 1.8 \pm 0.2 \times 10^{12}$~cm.

\citet{fav05} determined a magnetic field strength $B = 96$~G assuming pressure equilibrium. 
From Eq.~\ref{eqdeltai}, $B = 299 \pm 2$~G ($\Delta I / I = 0.05$ for this flare). This value is quite 
higher than that determined by \citet{fav05}. Thus, then Alfv\'en velocity inside the tube is 
$c_a = 6.5 \times 10^8$~cm\,s$^{-1}$ and the tube velocity $c_t = 1.8 \times 10^8$~cm\,s$^{-1}$. 
With this velocity, $L = 1.8 \pm 0.2 \times 10^{12}$~cm, as for the sonic wave.  This length is
compatible with the star-disk connection scenario. 

\begin{table*}
\caption{\label{table} Parameters of the flares.}
\centering
\small
\begin{tabular}{ccccccccccccc}
\hline\hline
      & \multicolumn{4}{c}{\citet{fav05}} & &  \multicolumn{2}{c}{\citet{get08a}} & & \multicolumn{3}{c}{This work} \\
\cline{2-5} \cline{7-8} \cline{10-12}
Src.& $T_\mathrm{peak}$ & $n_\mathrm{e}$ & $B^\mathrm{a}$ & 
                       $L$ & & $L$ & $R_\mathrm{cor}$ & $R_\mathrm{dust}^\mathrm{b}$ &  
                       $L_\mathrm{HD}$ & $L_\mathrm{MHD}$ & 
                       $B_\mathrm{min}$ & star-disk \\
      &  (MK) & ($10^{10}$ cm) & (G) & ($10^{12}$ cm) & & 
                       ($10^{12}$ cm) & ($10^{12}$ cm) & ($10^{12}$ cm) & ($10^{12}$ cm) & ($10^{12}$ cm) & (G) & connection \\ 
\hline
332   & 113 & 0.04 & 12   & 7.3                & & $0.5-1.7$     & ...     & 0.56 & $1.1 \pm 0.1$     & $0.98 \pm 0.11$  & $24 \pm 3$ & Y \\
597   &   87 & 5.40 & 128 & $0 - 0.75$    & & $0.08-0.47$  & ...     & 1.99 & $1.1 \pm 0.2$     & $0.90 \pm 0.09$  & $162 \pm 27$ & Y? \\
1246 & 270 & 5.60 & 240 & $0.32-0.47$ & & 0.24              & 0.54 & 0.48 & $0.34 \pm 0.04$ & $0.34 \pm 0.08$ & $365 \pm 45$ & N \\
1456 &  40\tablefootmark{c} & ... & ... & ... & & ... & ... & ... & $0.44 \pm 0.04$ & ... & ... & N? \\
1608 & 258 & 1.00 & 96   & $0-0.99$      & & $0.74-0.77$  & 0.60 & 0.75 & $1.8 \pm 0.2$ & $1.8 \pm 0.2$ & $299 \pm 2$ & Y \\
\hline
\end{tabular}
\tablefoot{$L_\mathrm{HD}$ and $L_\mathrm{MHD}$ are the loop lengths determined by us assuming the oscillation is 
an acoustic or a magneto-hydrodynamic wave, respectively.
\tablefoottext{a}{Magnetic field strength assuming pressure equilibrium \citep{fav05}.}
\tablefoottext{b}{Dust destruction radius from \citet{aar10}.}
\tablefoottext{c}{Value assumed in this work (see text).}
}
\end{table*}

\section{Summary and Conclusions}

We applied wavelet analysis tools to reveal oscillation patterns in the flare light curves of T Tauri stars observed 
by the COUP. 
With the results obtained from this analysis and the plasma characteristics previously determined, 
we inferred the length of the flaring loops. {Our analysis is restricted to those flares showing a significant 
oscillation. Other flares were also studied but their analysis gave uncertain results and they were not included here.} 
{The limitations of our results are linked only to the uncertainties in the plasma parameters taken 
from the literature. In particular, the plasma temperatures in \citet{fav05} are not well constrained in some 
cases.} {We interpret oscillations as caused by the fundamental mode and/or the first harmonic
inside the loop. Higher order harmonics require higher energies to be released \citep[e.g.][]{sel05}.}

{In our study, we did not assume the observed flares occurs in a single loop \emph{a priori}.
The wavelet analysis was applied to the light curves of flaring CTTS of the COUP. When 
an oscillation pattern was revealed, we determined plasma properties including the length of the 
oscillating structure. This value depends only on the oscillation period and on the wave velocity. In the 
very unlikely case of more than one loop oscillating in coherence (see Section~\ref{intro}), 
all of them would oscillate at the same frequency and the result would be applicable to each loop.} 

Assuming a simple physical model for triggering of an oscillation in a magnetic tube, other 
characteristics of the coronal medium can be inferred. In this work, we determined magnetic field 
strengths at the loop apex. 
Our results show that flares may take place in magnetic tubes connecting the star with its accretion disk. In 
particular, we find that at least three stars (COUP~332, COUP~597 --\, or V2252~Ori\,-- and COUP~1608 
--\,or OW~Ori\,--) have magnetic tubes {potentially} connecting the star with its accretion disk.  According to 
MHD simulations by \citet{orl11}, a flare in a magnetic tube connecting the star with the disk may trigger 
subsequent accretion by the star, in time scales of hours to days after the flare took place. In such a case, 
UV enhance should be detected in these stars after some hour after the flare maximum. A simultaneous 
long X-ray and UV observation of these stellar systems would eventually confirm this scenario. 

We conclude that analyzing oscillations in flare light curves is a powerful tool to 
investigate the actual extension of the corona of stars and determine other parameters. 
The particularly case of classical T Tauri stars is a good check point.


%
%

\end{document}